\documentclass{webofc}
\usepackage[varg]{txfonts}   
\usepackage{epstopdf}
\usepackage{wrapfig}
\begin{document}
\title{Hadron modification in a dense baryonic matter}
%
%

\author{\firstname{Genis} \lastname{Musulmanbekov}\inst{1}\fnsep\thanks{\email{genis@jinr.ru}}
}
\institute{JINR, Dubna}

\abstract{%
 Starting with the Strongly Correlated Quark Model of a hadron structure, SCQM, we demonstrate how the properties of mesons and baryons are modified in a hot and dense nuclear environment. These in-medium modifications can lead to the observable effects in heavy ion collisions, such as enhancement of strangeness and dropping vector meson masses.
}
\maketitle
\section{Introduction}
\label{intro}
Current and future experiments focus on observables which are sensitive to QGP phase transition, especially to the range of the phase diagram which close the critical point. Observables indicating non-monotonic and unexpected (from theoretical point of view) behavior of emitted particles are particularly important.
\begin{wrapfigure}{r}{0.4\textwidth}
\centering
\includegraphics[width=0.4\textwidth]{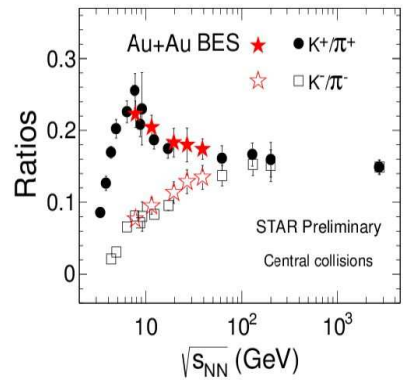}
\vspace{-20pt}
\caption{Energy dependence of $K^{\pm}/\pi^{\pm}$ ratio for central heavy ion collisions at midrapidity.}
\vspace{-10pt}
\label{horn}
\end{wrapfigure}
In this way the study of the strange particle production in heavy ion collisions is promising as they could serve a good diagnostic tool to investigate the properties of nuclear matter under extreme conditions.
The systematic study of hadron production in central Pb+Pb collisions at SPS performed by NA49 collaboration revealed a sharp structure in energy dependence of positive kaon to pion multiplicity ratio, $K^{+}/\pi^{+}$ \cite{horn-SPS}.
That peculiarity, called "horn"--effect, was later confirmed by Beam Energy Scan (BES) program of STAR collaboration at RHIC \cite{horn-STAR}  (Fig. \ref{horn}). At the same time there were no any peculiarities observed in the energetic behaviour of $K^{-}/\pi^{-}$ ratio.
The idea that strangeness is a good signal of deconfinement was put forward by J. Rafelski in 1982 \cite{Rafelski}.  The argument was the following: it is energetically favourable to produce $s\overline{s}$ - pairs in deconfined medium than a pairs strange hadrons in hadron gas. Interpretation of the non--monotonic structure of $K^{+}/\pi^{+}$ has initiated intense theoretical activity. Authors attempted to reproduce the horn structure employing approaches either with phase transition to QGP or without it. "Horn" -- like structure has been predicted in Ref. \cite{gasd}, as a manifestation of phase transition between thermalized hadronic and partonic phases. Albeit a variety of models, statistical \cite{hrgm1, hrgm2, hrgm3, hrg-Hag} and kinetic \cite{nayak, tomas} (with or without deconfinement) have been proposed for interpretation of "horn" structure its satisfactory understanding is still not complete.
\begin{wrapfigure}{r}{0.4\textwidth}
\centering
\includegraphics[width=0.4\textwidth]{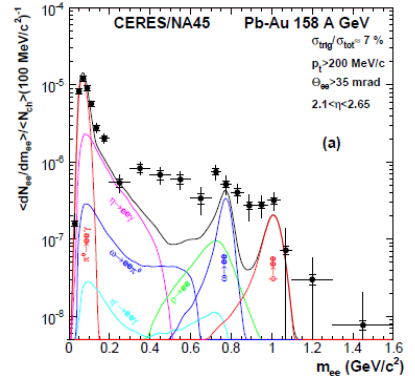}
\vspace{-20pt}
\caption{Invariant-mass spectrum of $e^{+}e^{-}$--pairs compared to the expectation from the hadron decay cocktail \cite{di2}.}
\vspace{-10pt}
\label{dilep}
\end{wrapfigure}

Another promising observable is a yield of dileptons. Dileptons are an ideal probe to study the properties of hot and dense nuclear matter, since they are emitted at different stages of reaction and escape the medium nearly unperturbed. They allow unique access to the properties both of the medium and resonances that decay within a strongly interacting medium. Measurements of emission of dielectrons  in different nuclear reactions at wide range of collision energy revealed an enhancement of invariant mass spectra of di-leptons yield in the interval 0.2 - 0.6 GeV \cite{di1, di2, di3} (Fig. \ref{dilep}). This enhancement was interpreted as in-medium modifications of hadrons at high temperature and density resulting in strong broadening of the  $\rho$--meson and/or its``mass--dropping'' \cite{leo, hay, bratkov}.
We propose our interpretation of the observed phenomena using for this purpose the Strongly Correlated Quark Model, SCQM, developed by the author \cite{Mus0}.
\section{The model}
\label{sec-1}
The real physical vacuum, the energy of which is below the ``empty'' perturbative vacuum,  is populated by gluon and quark--antiquark condensates. Imagine hypothetically a single quark of a certain color embedded in the physical vacuum. The color field of the quark polarizes the surrounding vacuum creating a condensate.  At the same time it experiences the pressure of the vacuum, as a reaction on the ordering, because of the presence of quantum fluctuations of gluon and quark--antiquark fields, or zero point radiation field in a classical sense. Suppose we place a corresponding antiquark in the vicinity of the first quark. Owing to their opposite signs, color polarization fields of the quark and antiquark interfere destructively in the overlap regions eliminating each other maximally at the
middle-point between them. This effect leads to a decreasing value
of the condensate density in that region and overbalancing of
the isotropic vacuum pressure acting on the quark and antiquark. As a result, an attractive force between the quark and antiquark emerges and the quark and antiquark start to move
towards each other. The density of the remaining condensate around
the quark (antiquark) is identified with the hadronic matter
distribution which is associated with a dynamical mass of the quark. At maximum displacement in the $\overline{q}q$ system
corresponding to small overlap of color fields, hadronic
matter distributions have maximum extent and densities. The quark (antiquark) in this state possesses a constituent mass. The closer
they come each other, the larger is the destructive interference
effect and the smaller hadronic matter distributions around quarks
and the larger their kinetic energies. In this state the quark (antiquark) becomes relativistic with a current mass. So, the quark and
antiquark start to oscillate around their middle-point. For such
interacting $\overline{q}q$ pair located from each other on a
distance $2x$, the total Hamiltonian is

\begin{equation}
H=\frac{m_{\overline{q}}}{(1-\beta ^{2})^{1/2}}+\frac{m_{q}}{(1-\beta
^{2})^{1/2}}+V_{\overline{q}q}(2x),
\end{equation}%
where $m_{\overline{q}}$, $m_{q}$ are the current masses of the valence
antiquark and quark, $\beta =\beta (x)$ is their velocity depending on
displacement $x$, and $V_{\overline{q}q}$ is the quark--antiquark potential
energy with separation $2x.$ It can be rewritten as
\begin{equation}
H=\left[ \frac{m_{\overline{q}}}{(1-\beta ^{2})^{1/2}}+U(x)\right] +\left[
\frac{m_{q}}{(1-\beta ^{2})^{1/2}}+U(x)\right] =H_{\overline{q}}+H_{q},
\label{hamil}
\end{equation}%
where $U(x)=\frac{1}{2}V_{\overline{q}q}(2x)$ is the potential
energy of the quark or antiquark. We postulate that the potential energy of quark is equal to its dynamical mass:
\begin{equation}
2U(x)=\int_{-\infty }^{\infty }dz^{\prime }\int_{-\infty }^{\infty
}dy^{\prime }\int_{-\infty }^{\infty }dx^{\prime }\rho (x,{\mathbf{r}%
^{\prime }})\approx 2M_{Q}(x)
\label{poten-mass}
\end{equation}%
with
\begin{equation}
\rho (x,\mathbf{r}^{\prime })=c\left| \varphi (x,\mathbf{r}^{\prime })\right|
^{2}=c\left| \varphi _{Q}(x^{\prime }+x,y^{\prime },z^{\prime })-\varphi _{%
\overline{Q}}(x^{\prime }-x,y^{\prime },z^{\prime })\right| ^{2}.
\label{color-field}
\end{equation}
where $\rho$ is the resulting density of hadronic matter (quark-antiquark condensate) formed by color fields $\varphi$ and $\varphi _{\overline{Q}}$ of the quark and antiquark, respectively. The the structure and shape of vacuum polarization around the color quark/antiquark which could give us the information about the confining potential is not known.

It turnes out that our quark--antiquark system behaves similarly to the breather solution of one--dimensional Sine-Gordon equation \cite{Raja} which in scaled form reads
\begin{equation}
\Box \phi (x,t)+\sin \phi (x,t)=0,
\end{equation}
where $\phi(x,t)$ is a scalar function and $x$ and $t$ are dimensionless.
It has a so-called {\it breather} solution
\begin{equation}
\phi(x,t)_{br}=4\arctan \left(\frac{\sqrt{1-w^{2}}\sin(wt)}{w\cosh(\sqrt{1-w^{2}x})}\right),
\end{equation}
which is the periodic soliton--antisoliton solution for frequencies $w<1$.  The energy density profile of the soliton--antisoliton system
\begin{equation}
\varphi(x,t)_{br}=d\phi(x,t)_{br}/dx
\end{equation}
oscillate the same way as our quark--antiquark system. W. Troost \cite{troost} demonstrated that the Hamiltonian (\ref{hamil}) corresponds to the breather (soliton--antisoliton) solution of Sine-Gornon equation.  He derived the effective potential $U(x)$ for this solution
\begin{equation}
U(x)=M\tanh ^{2}(\alpha x),
\label{poten}
\end{equation}
where $M$ is a mass of soliton/antisoliton and $\alpha$ is an adjusting parameter. Hence, we can identify our potential of quark--antiquark interaction in hamiltonian (\ref{hamil}) with the potential of soliton--antisoliton interaction.

\begin{figure}[h]
\vspace{-50pt}
\centering
\includegraphics[width=4.5in]{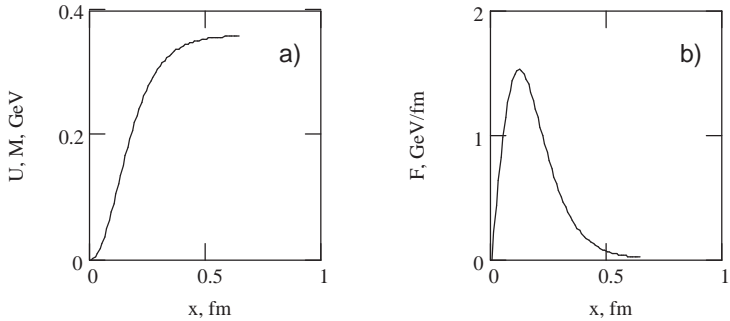}
\vspace{-40pt}
\caption{a) Potential energy of a quark and its dynamical (constituent) mass versus its displacement from the origin of oscillation; b) ''Confinement'' force.}
\vspace{-10pt}
\label{pot-force}
\end{figure}

Since quarks are the members of the fundamental color triplet,
generalization to the 3-quark system (baryons, composed of Red,
Green and Blue quarks) is performed according to
$SU(3)_{color}$ symmetry: a pair of quarks has coupled
representations $3\otimes 3=6\oplus \overline{3}$ and for quarks
within the same baryon only the $\overline{3}$ (antisymmetric)
representation is realized. Hence, an antiquark can be replaced by
two correspondingly colored quarks to get a color singlet baryon;
destructive interference takes place between color fields of three
valence quarks (VQs). Putting aside the mass and charge differences
of valence quarks one can consider three quarks
oscillating synchronously along the bisectors of equilateral triangle turning from the constituent to current state and inversely.
Therefore, the model unifies the features of bag models and constituent models. At a maximal displacement quark becomes nonrelativistic with constituent mass corresponding to the maximal value of condensate surrounding it. Further, owing to the prevailing condensate pressure from the outside, it moves under influence of the potential (\ref{poten}) (see Fig. \ref{pot-force}a) towards two other quarks, and at the origin of oscillation it becomes relativistic with the current mass. Thus, during oscillation quarks transit from constituent states to current states that corresponds to dynamical chiral symmetry braking and restoration. Important feature of the model is that there is no a confining potential/force inside a nucleon.
During oscillations (putting aside Coulomb and spin interactions) the interaction force between quarks vanishes both at the origin of oscillation and at a maximal displacement (Fig. \ref{pot-force}b). It becomes maximal in between the origin and maximal displacement. Thus, at the origin of oscillations quark and antiquark in mesons and three quarks in baryons do not interact, i.e. they are in the state of asymptotic freedom. As to real confining potential, it should act at distances exceeding hadronic radii. Apparently, ``imprisonment'' of quarks is a consequence of the topological nature of hadrons. Hereinafter we assume that the quark--antiquark describing mesons and three quark systems describing baryons are topological solitons. Topological solitons are characterized by the conserving numbers, so--called, winding numbers. For baryons a winding number is identified with the baryonic number.  It means that at any temperature and density of nuclear environment baryon conserves its identity and baryonic number.
The model meets the local gauge invariance. Indeed, suppose $\psi_{color}$ is a wave function of a single quark in color space where index {\it color} accepts one of the values Red, Green, Blue. Interactions of of R, G, and B quarks in a nucleon which result in their oscillations can be reduced to the phase rotation the wave function $\psi_{color}$ of each quark in color space
\begin{equation}
\psi_{color}(x) \rightarrow e^{i\theta(x)}\psi_{color}(x).
\label{psi_col}
\end{equation}
This phase rotation results in dressing (undressing) of the quark by quark/qluon condensate that can be linked with transformation of the gauge field $A^{\mu}$:
\begin{equation}
A^{\mu}(x) \rightarrow A^{\mu}(x)+\partial^{\mu}\theta(x).
\label{gauge-eq}
\end{equation}
Here we dropped color indices $A^{\mu}_{a}(x) \rightarrow A^{\mu}(x)$ since interactions of color quarks via non--Abelian fields of QCD in our model are reduced to its electrodynamical analog
\begin{equation}
F^{\mu\nu}_{a}=\partial^{\mu}A^{\nu}_{a}-\partial^{\nu}A^{\mu}_{a}-\lambda f^{abc}A^{\mu}_{b}A^{\nu}_{c} \rightarrow F^{\mu\nu}=\partial^{\mu}A^{\nu}-\partial^{\nu}A^{\mu}.
\label{Field}
\end{equation}

The parameters of the model are the maximum displacement of valence quark and antiquark in mesons and 3 quarks in baryons, $x_{\max }$, and the parameters of the hadronic matter distribution formed by quark--antiquark condensate around them. In the absence of knowledge about the shape of quark--antiquark condensate around valence quarks, or the form of hadronic matter in a constituent quark $\varphi_{Q(\overline{Q})}$, we take it in a gaussian form
\begin{equation}
\varphi _{Q(\overline{Q})}(x,y,z)=\varphi _{Q(\overline{Q})}(x_{1},x_{2},x_{3})=%
\frac{(\det \hat{A})^{1/2}}{(\pi )^{3/2}}\exp \left( -\mathbf{X}^{T}\hat{A}%
\mathbf{X}\right) ,
\label{gaussian}
\end{equation}
where the exponent is written in a quadratic form.

The value of the maximal quark (antiquark) displacement,  and parameters of the gaussian function for hadronic matter distribution around VQ are chosen to
be $x_{\max }=0.64\ ${fm}$,\ \sigma _{x,y}=0.24\ ${fm}$,\ \sigma
_{z}=0.12\ ${fm}$.$ They are adjusted by comparison of calculated
and experimental values of the total, inelastic and
differential cross sections for $pp$ and $\overline{p}p$ collisions \cite%
{Mus2}. The mass of the constituent quark at maximum displacement is taken as $%
M_{Q(\overline{Q})}(x_{\max })=\frac{1}{3}\left( \frac{m_{\Delta }+m_{N}}{2}%
\right) \approx 360\ ${MeV}$,$ where $m_{\Delta }$ and $m_{N}$ are
masses of the delta isobar and nucleon correspondingly. The current
mass of the valence quark is taken to be $5\ ${MeV}.
\section{Hadron properties in heavy ion collisions}
\label{sec-2}
\begin{wrapfigure}{r}{0.4\textwidth}
\vspace{-10pt}
\centering
\includegraphics[width=0.4\textwidth]{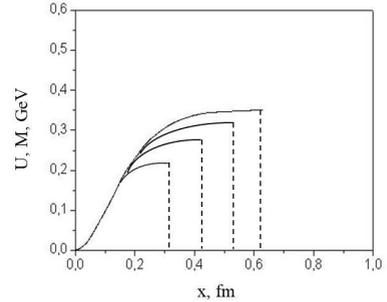}
\vspace{-20pt}
\caption{Modification of mass of constituent quark inside compressed nuclear matter.}
\vspace{-10pt}
\label{poten-mod}
\end{wrapfigure}

In head--on collisions of two heavy ion nuclei the energy density in the overlap zone increases drastically. The time of ``crossing'' two symmetric nuclei through each other when they cease to overlap is $t_{cross}=2R/\gamma$, where $R$ is the rest--frame radius of the nucleus. Excited baryons and secondaries created in the overlap zone can be considered ``formed'' at some proper time $\tau_{form}$ which is $\sim$ 1 fm/c. At low and moderate collision energies where $\tau_{form} < t_{cross}$ particle production and their interactions take place mainly in the overlap zone with high baryonic density. At very high collision energies, once the remnants of Lorenz--contracted disks recede after their initial overlap, the region between them is occupied by a hot and dense ``fireball'' of interacting secondaries characterized by low baryonic density.  The general point of view claims that the hadronic matter at these conditions undergoes the phase transition to quark gluon plasma where quarks become deconfined and the chiral symmetry is restored.
Being based on the above model of nucleon structure we offer other scenario. We start with low and intermediate collision energies, when $\tau_{form} < t_{cross}$. In the initial stage of collision of nuclei, due to propagation through each other and their Lorenz contraction, the baryonic density in the overlap zone increases more than twice. Correspondingly, the accessible volume occupied by each nucleon composed of light quarks is reduced, at least, more than twice.  As a result of accessible volume reduction, the vacuum condensate around the valence quarks decreases that, in turn, leads to reduction of the dynamic mass of quarks and amplitude of oscillation, as shown in Fig. \ref{poten-mod}. Further compression of nuclear matter could lead to a collapse of nucleons. To avoid collapsing it is preferable to nucleons to transit to delta--isobars and their excited states: $p, n \rightarrow \triangle, \triangle^{\ast}$... Parallel alignment of spins of all three quarks leads to their repulsion (according to Pauli principle) that could prevent the collapsing process. Therefore, there should be a limit of accessible volume reduction which can be specified as a ``hard-core'' of delta isobars and their excited states.
 However, at higher compression this mechanism is not sufficient because the cores of light quarks need to occupy relatively large volume. Moreover, at higher compression it is preferable for  nucleons to be converted to hyperons, as their dimensions/cores are small compared with cores of deltas. This transition of nucleons to hyperons can be described in the framework of $^{0}P_{3}-$model of vacuum. In a compression zone the production of $s\overline{s}-$pairs should be dominating in the content of condensate. $s-$quarks of the pairs replace one or more of $d/u-$quarks of the nucleons, and $\overline{s}-$quarks form with those replaced quarks strange mesons: $p,n\longrightarrow hyperons+kaons$.
In these transition channels the $K^{+}$s and $K^{0}$s, but any $K^{-}$ are produced only \cite{Mus4}. At higher compression the production of heavier resonances with all three quark spins aligned parallel should be dominating.  However, the transition mechanism works if the ``crossing'' time, $t_{cross}$, is larger than formation time, $\tau_{form}$. With increasing collision energy $t_{cross}$ becomes very short, and the Lorenz-contracted disks with excited baryons fly away leaving behind the hot and dense fireball with a low baryonic chemical potential. Hence, at $\tau_{form} > t_{cross}$ the transition mechanism ceases to work. As demonstrated in \cite{Mus4}, this mechanism can result in the non--monotonic behavior of the $K^{+}/\pi^{+}-$ratio, the "horn"--effect observed in the experiments. Obviously, this mechanism is additional to the particle production while nuclei propagate through each other. Among produced particles, according to the above arguments, the production of baryon resonances and vector mesons should be dominating.

At essentially high energies the flying away excited remnants of colliding nuclei leave behind them a highly compressed fireball with a low composition of baryons. Its evolution starting with multiparticle production leads to its heating and thermalization and at the last stages to its expansion. Again, since the nuclear matter inside the fireball is highly compressed the production of pions composed of light $u$ and $d$ quarks in pseudoscalar state is suppressed, and vector mesons, $\rho, \omega, \phi$, and heavier mass resonances will be dominating in the composition of fireball. At the same time there should be essential modification of the features of mesons composed of light quarks ($\rho, \omega$) in a compressed medium. Because of reduction of available volume and, correspondingly, decreasing of condensates around quarks, the masses of these mesons will be depending on the compression value or medium density (Fig. \ref{poten-mod}). What follows from our model,  the more particle density is inside the fireball, the less is the mass of (vector) mesons produced. Without knowing the parameters of fireball we can be express this dependence as
\begin{equation}
m^{*}=m_{0}(1-\alpha \rho/\rho_{0})^{\beta},
\label{mass-drop}
\end{equation}
where $\alpha$ and $\beta$ are adjustable parameters. In hadronic channels the vector mesons can decay up to the threshold, 2$m_{\pi}$. In dilepton decay mode the threshould continues down to 2$m_{e}$. Therefore, in the framework of our approach, the enhancement of spectral functions of vector mesons (Fig. \ref{dilep}) can be explained by domination of their  production and mass dropping. Moreover, the mass dropping effect can be accompanied by the resonance decay width dependence
\begin{equation}
\Gamma_{R}\sim\Gamma_{R}^{0} (\rho/\rho_{0})^{\gamma},
\label{width-broad}
\end{equation}
which results in increasing lifetime of resonances $\tau=1/\Gamma_{R}^{0}$. Both effects leads to suppression of multiparticle production in a hot and dense medium.  During the expansion of fireball, accompanied simultaneously by its cooling, the physical vacuum inside it is restored, that leads to restoration of hadron features.

Analyzing the "horn"-effect and the enhancement of invariant mass spectra of dielectrons in the framework of proposed model, SCQM, we demonstrate that baryons and mesons in a hot and dense medium are essentially modified.


\end{document}